\documentclass[runningheads]{llncs}
\usepackage{graphicx}

\newif\ifanonymous
\newif\ifcameraready

\anonymousfalse

\camerareadytrue

\usepackage{hyperref}

\usepackage{tabularx}

\usepackage{ifthen} 

\newboolean{auxinfo}
\setboolean{auxinfo}{true}   

\newboolean{showoutline}

\setboolean{showoutline}{false}

\newboolean{showcomment}

\setboolean{showcomment}{true} 

\usepackage{ifthen}

\ifthenelse{\boolean{auxinfo}}
  {}
  {
    \setboolean{showoutline}{false}
    \setboolean{showcomment}{false}
  }

\usepackage{savesym}
\usepackage{xcolor}
\savesymbol{Bbbk}
\usepackage{amssymb}
\restoresymbol{AMS}{Bbbk} 

\ifthenelse{\boolean{showcomment}}
   {\newcommand{\nb}[2]{\fcolorbox{gray}{yellow}{\bfseries\sffamily\scriptsize#1}{\sf\small$\blacktriangleright${\em #2}$\blacktriangleleft$}}
   \newcommand{\working}[1]{\fcolorbox{gray}{yellow}{{\bf #1}\emph{\scriptsize---in progress---}}}
   \newcommand{\TBD}[1]{\fcolorbox{gray}{yellow}{{\bf #1}\textbf{TBD}}} 
  }
  {\newcommand{\nb}[2]{}{}
   \newcommand{\working}[1]{}
   \newcommand{\TBD}[1]{} 
  }

\usepackage{todonotes}
\ifthenelse{\boolean{showoutline}}{
	\newcommand{\outline}[3]{
		~\newline 
		\fcolorbox{red}{white}{
			\parbox{\columnwidth-13.465pt}{
				\ifthenelse{\equal{#1}{}}{
					\ifthenelse{\equal{#2}{}}{
						\noindent\colorbox[rgb]{0.65,0.16,0}{\textcolor[rgb]{1,1,1}{\textbf{Outline}}}
					}{
						\colorbox[rgb]{0.65,0.16,0}{\textcolor[rgb]{1,1,1}{\textbf{Outline -- Responsible: #2}}}
					}
				}{
					\ifthenelse{\equal{#2}{}}{
						\noindent\colorbox[rgb]{0.65,0.16,0}{\textcolor[rgb]{1,1,1}{\textbf{#1 page(s)}}}
					}{
						\colorbox[rgb]{0.65,0.16,0}{\textcolor[rgb]{1,1,1}{\textbf{#1 page(s) -- Responsible: #2}}}
					}
				}
				#3
			}
		}
	}
}{
	\newcommand{\outline}[3]{}
}

\newcommand\defauxcomm[1]{
       \expandafter\newcommand\csname #1NB\endcsname[1]{\nb{#1}{##1}}
       \expandafter\newcommand\csname #1WK\endcsname{\working{#1}}
       \expandafter\newcommand\csname #1TBD\endcsname{\nb{#1}}
    } 

\defauxcomm{ks} 
\defauxcomm{he} 
\defauxcomm{se} 
\defauxcomm{ck} 
\defauxcomm{lg} 
\defauxcomm{ak} 
\defauxcomm{lf} 
\defauxcomm{kl} 
\defauxcomm{cs} 
\defauxcomm{aa} 
\defauxcomm{tw} 
\defauxcomm{ag} 
\defauxcomm{fb} 
\defauxcomm{lm} 

\defauxcomm{mn}
\defauxcomm{fk}
\defauxcomm{ek}

\usepackage[normalem]{ulem}
\ifthenelse{\boolean{showcomment}}
    {\newcommand{\strike}[1]{\textcolor{red}{\sout{#1}}}}
    {\newcommand{\strike}[1]{}}

\newcommand{\tag}[1]{{\footnotesize\texttt{[#1]}}}

\begin{document}
\title{Towards an MLOps Architecture for XAI in Industrial Applications}

%
%

\ifcameraready
\author{Leonhard~Faubel\inst{1} \and
Thomas~Woudsma\inst{2} \and
Leila~Methnani\inst{3} \and
Amir~Ghorbani~Ghezeljhemeidan\inst{4} \and
Fabian~Buelow\inst{5} \and
Klaus~Schmid\inst{1} \and
Willem~D.~van~Driel\inst{4, 6} \and
Benjamin~Kloepper\inst{5, 7} \and
Andreas~Theodorou\inst{3} \and
Mohsen~Nosratinia\inst{8} \and
Magnus~B\r{a}ng\inst{9}
}
\fi

\ifanonymous
\author{Anonymous}
\fi

%
\ifcameraready
\authorrunning{L. Faubel et al.}
\fi
\ifanonymous
\authorrunning{Anonymous}
\fi
%
\ifcameraready
\institute{University of Hildesheim, Universitätsplatz 1, 31141 Hildesheim 
\email{\{faubel,schmid\}@sse.uni-hildesheim.de}\\
\and
Prodrive Technologies, Science Park Eindhoven 5501, 5692 EM SON 
\and
Umeå University, universitetstorget 4, 901 87 Umeå 
\and
Delft University of Technology, Mekelweg 5, 2628 CD Delft 
\and
ABB Corporate Research Germany, Wallstadter Str. 59, 68526 Ladenburg 
\and
Signify High Tech Campus 48, 5656 AE Eindhoven 
\and
Capgemini Invent, Mainzer Landstrasse 178-190, 60327 Frankfurt am Main 
\and
Viking Analytics, Anders Carlssons gata 14, 417 55 G\"oteborg 
\and
Linköping University, 581 83 Linköping 
%
}
\fi
\ifanonymous
\institute{Anonymous}
\fi
\maketitle              
\begin{abstract}

Machine learning (ML) has become a popular tool in the industrial sector as it helps to improve operations, increase efficiency, and reduce costs. However, deploying and managing ML models in production environments can be complex. 
This is where Machine Learning Operations (MLOps) comes in. MLOps aims to streamline this deployment and management process. 
One of the remaining MLOps challenges is the need for explanations. These explanations are essential for understanding how ML models reason, which is key to trust and acceptance. Better identification of errors and improved model accuracy are only two resulting advantages. 
An often neglected fact is that deployed models are bypassed in practice when accuracy and especially explainability do not meet user expectations. We developed a novel MLOps software architecture to address the challenge of integrating explanations and feedback capabilities into the ML development and deployment processes. In the project \ifcameraready\textit{EXPLAIN}\fi\ifanonymous Anonymous\fi, our architecture is implemented in a series of industrial use cases. 
The proposed MLOps software architecture has several advantages. It provides an efficient way to manage ML models in production environments. Further, it allows for integrating explanations into the development and deployment processes. %

\end{abstract}
\section{Introduction}
\label{sec:introduction}





The application of ML in the industrial sector promises significant improvements such as increased effectiveness, energy efficiency, and yield. However, despite many pilot applications, the practitioners among the authors observe that only a few ML projects have moved into actual and continuous production use. 
One of the barriers to the successful and sustainable use of ML is the difficulty in communicating the inferences, predictions, and decisions these algorithms make to domain experts who may not have a technical background in ML. 
Such communication cannot be limited to the output of the ML model but must also include insights into how and why the model produced that output. This \emph{explanation} is necessary to create trust and enable domain experts to exercise oversight over both the ML development process and the ML models in use. 

The \ifcameraready ITEA project EXPLAIN~\cite{EXPLAIN_homepage} \fi \ifanonymous Anonymous Project \fi aims to develop an end-to-end ML life cycle and an MLOps software architecture that inherently provides explainability and interactivity for industrial domain experts. This means that individuals with little to no technical background in ML can participate and contribute during the entire process, during activities like data preparation, modeling, model deployment, and inference. The process becomes accessible and transparent so that everyone involved can understand how and why a model generates its output and even interact with it to contribute to human domain knowledge. 

To date, there has been limited use and discussion of MLOps for XAI in industrial applications. If at all, it has been used only for specific elements or individual steps in the ML life cycle. As we learned from an internal study~\cite{swora_investigation_2023}, major cloud providers now offer such integrated MLOps solutions, but only some currently include specific XAI-based functionalities and the specific needs of industrial applications are not considered. 
In this paper, we propose an MLOps architecture with the above-mentioned capabilities. This architecture is based on our companies' experiences and project requirements. We believe this architecture will help bridge the gap between technical and non-technical experts and pave the way for more transparent and accessible ML processes. 

At first, the problem is described in more detail using the project life cycle in Chapter~\ref{sec:life-cycle}. Then, the related work in MLOps, MLOps workflow, XAI, and interactive ML is described in Chapter~\ref{sec:related_work}. Chapter~\ref{sec:requirements} summarizes our architecture's MLOps and XAI requirements. 
Based on these requirements, Chapter~\ref{sec:results} briefly describes the novel software architecture. The Implementation is discussed in Chapter~\ref{sec:discussion}. Finally, Section~\ref{sec:conclusion} concludes.

\section{Explain Life Cycle}
\label{sec:life-cycle}

\begin{figure}[h]
  \centering
  \includegraphics[clip=true, trim=0cm 2cm 3cm 1.5cm, width=1\textwidth]{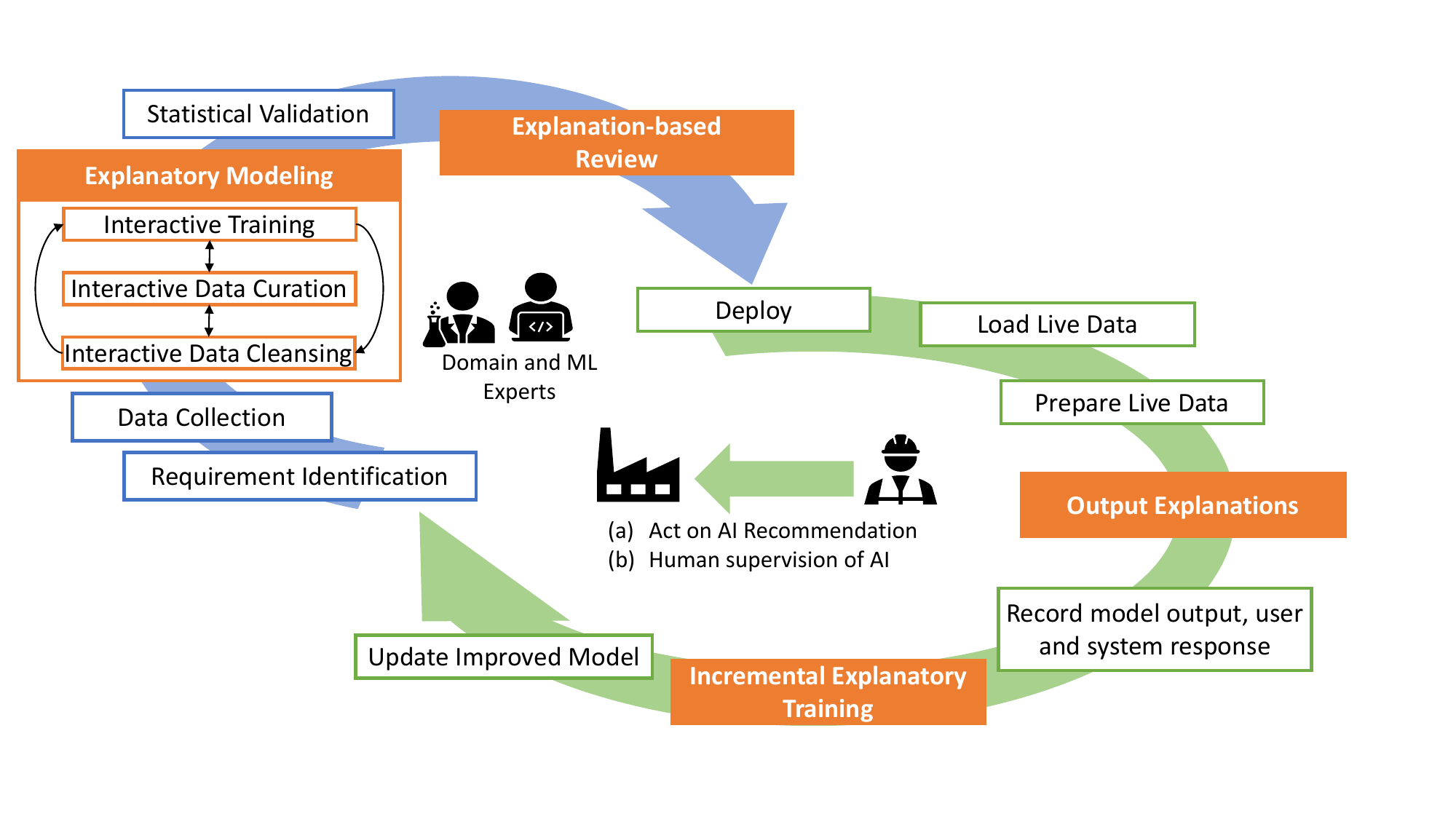}
  \caption{\label{fig:explain_MLOps_lifecycle} Adopted explain life cycle from \cite{project_proposal}.}
\end{figure}

We aim to enhance the traditional ML life cycle by adding steps that empower stakeholders and elevate their influence, participation, and ownership. This vision is driven by a desire to create a more practice-integrated ML approach involving practitioners in the process and outcome. By engaging stakeholders at every stage, \ifcameraready EXPLAIN \fi \ifanonymous Anonymous \fi seeks to create a more transparent and accountable ML process that delivers better results and benefits for all. The extended life cycle is shown in Figure~\ref{fig:explain_MLOps_lifecycle}.

As part of the ML life cycle and MLOps, addressing the detection and update of ML models whose inference quality is deteriorating is vital. A common reason for this quality degradation is concept drift. Concept drift can occur when the modeled relation evolves in such a way that the data model (a concept a model has learned during training) deteriorates with progressing time. For example, the gradual degradation of physical process equipment, such as fouling in a heat exchanger or wear of a pump, can cause concept drift for models used in a production environment. Closely related to concept drift is data drift, which is one of the main reasons for deteriorating model quality. Data drift can be described as a change in the distribution of model input data during production with respect to the distribution of input data during model training. Predictions can worsen with progressing data drift if the model does not account for it. An example of data drift is a change of input material quality in a process, e.g., due to a change of material supplier or sensor drift influencing a model feature.\\
Various techniques detect data drift and potential concept drift, e.g.,~\cite{barros2018large,wang2015concept,zenisek2019machine,yu2019concept}. These techniques usually rely on statistical tests or additional data models. Some techniques provide a certain degree of explainability to the drift detection process~\cite{adams2023explainable} and others rely on explainability methods to detect drift~\cite{duckworth2021using}.\\
Explanatory training of models based on feedback from domain experts on the explanations generated by the model itself can complement current approaches~\cite{Mittelstadt2019Jan}. The explanations can also be used as plausibility checks for the models and their underlying data sets, helping to avoid technical debt in the ML lifecycle and in the ongoing development and deployment of ML/AI applications~\cite{Sculley2015,BarredoArrieta2020Jun,TD_IND}. 
Appropriate stakeholders are described in Section~\ref{sec:stakeholders} while the life-cycle steps are described in Section~\ref{sec:life-cycle-steps}. 


\subsection{Stakeholders}
\label{sec:stakeholders}
For explanations to be useful, the receiver of any particular explanation should be carefully and purposefully considered~\cite{BarredoArrieta2020Jun}. For instance, a direct user of an AI system may ask a ``why not'' question, seeking clarification for why their expectations were not met, while a system engineer may ask a ``how'' question, intending to understand where to debug and improve the overall system performance. Therefore, it is crucial first to identify \textit{who} our various stakeholders are and further pinpoint their corresponding requirements for explainability at any given time in the life cycle~\cite{kotriwala2021xai}. 

In the project \ifcameraready EXPLAIN\fi \ifanonymous Anonymous\fi, we involve industry stakeholders throughout the design and development of our architecture to meet their needs. In addition to the ML engineers and data scientists who work on training, deploying, and maintaining the ML models~\cite{faubel_interviews}, we identify two main stakeholder groups: end users and domain experts. Our ML life cycle, as illustrated in Figure~\ref{fig:explain_MLOps_lifecycle}, heavily relies on these stakeholders. End users like machinery operators, site supervisors, and machine maintenance personnel are considered, as well as domain experts such as process engineers and reliability experts. The specific stakeholders involved in each stage of the ML life cycle depend on the use case domain, which in our case, includes industries like \textit{manufacturing}, \textit{electronics}, \textit{mining}, and \textit{pulp and paper}.

Consider the pulp and paper domain, where manufacturing involves continuous maintenance and quality control of large machinery, which can be optimized using smart sensors. These sensors are installed on rotational parts of large manufacturing machines to detect and collect vibration data for further predictive analytics. Vibration analysts are one of the stakeholders requiring explainability---these are the aforementioned domain experts who support predictive model development. Stakeholders with a deep understanding of the critical variables may want to examine a specific prediction and inquire about the factors that led to that outcome. The XAI technique to look towards an appropriate answer may be \textit{feature-importance}, where the most influential features over that particular prediction are offered in order of importance. If a feature contributes too heavily---or not at all---towards the prediction, then the domain expert can identify the issue and work with the ML expert to address it during model improvement. 
After ML model deployment, the pulp mill operators interact with the model; they receive recommended actions, which they may accept or reject. These operators are the end users, who also play a central role in our overall XAI process. 
As operators are individuals in specific roles, they make decisions on a day-to-day basis and their experience in that role heavily influences their choices, regardless of any AI-driven tools available to them. It can be challenging to capture and represent this experience in ML-driven decision-making, which is why explainability is of interest to many. In particular, users may want to understand why a specific action was recommended over what they would have preferred. In these situations, counterfactual explanations may be helpful, as they can answer why a certain outcome was predicted instead of the one the user expected \cite{wachter2017counterfactual}. By presenting a minimal change in the input that would result in an alternative target outcome, counterfactual explanations can help users understand the reasoning behind the AI-powered recommendation. 

\subsection{Life-Cycle Steps}
\label{sec:life-cycle-steps}

The \textbf{blue circle} in Figure~\ref{fig:explain_MLOps_lifecycle} describes the initial development of an ML model for industrial applications. The industrial process is also the primary data source for historical data for ML training and new data for ML predictions, which are either processed directly by the AI systems or by the end user based on the recommendations of an AI system. In an industrial context, this initial process cannot be carried out by ML experts alone, but it is essential to involve professionals with a deep understanding of the industrial process. Ideally, these stakeholders will provide input on today's model requirements and support data collection and processing~\cite{project_proposal}. The orange boxes indicate process steps that allow for better involvement of industry experts and end-users. \emph{Explanatory modeling} combines interactive ML and explainable AI (XAI). Similar to active learning, \cite{settles2009active} ML models are incrementally trained as domain or ML experts improve the training data. Such improvements can take the form of sample labeling, training set curation (removing harmful samples or up-sampling beneficial samples), or data sample cleaning. As part of the process, the experts receive model outputs together with explanations of the model and provide feedback - which can be used to improve the training data set \cite{Teso2019Jan} or to fine-tune the loss function \cite{Schramowski2020Aug}. In the \emph{explanation-based review} phase, ML models are validated by providing domain experts with insight into the internal reasoning of the trained model. This ensures that the models learn only relevant concepts from the data provided and that only robust and reliable ML models are released for use in production.

After deployment, the ML applications enter the production phase, indicated with the \textbf{green circle}, acting on live production data. In this phase, \emph{output explanations} provide the end user with insights into why and how the model produced a particular output. This enables the end user to monitor the ML model and analyze the output. The explanation can also help to understand problems in production processing and to derive the right corrective actions more quickly by pointing the end user to relevant data points.  End users can provide feedback or trigger \emph{incremental explanatory training}.

\section{Related Work}
\label{sec:related_work}

Numerous publications cover MLOps, Explainable AI (XAI), and the MLOps workflow. However, no publications specifically focus on MLOps software architecture supporting explanations. 
Section~\ref{MLOps} defines MLOps and MLOps architecture in the industrial context, and Section~\ref{sec:related_word_mlops_workflow} summarizes essential sources on  the MLOps workflow. Section \ref{related_XAI} deals with XAI, while Section~\ref{sec:XIL} deals with interactive ML. 

\subsection{MLOps}
\label{MLOps}

The concept of DevOps~\cite{Sanjeev2017}, which pertains to the development and operation of expansive software systems, has gained significant popularity intending to accelerate the deployment and ensure reliable releases. MLOps is an evolving discipline focused on efficiently deploying and managing ML models in production environments. It combines the principles of DevOps with the specific challenges and requirements of ML systems, allowing organizations to operationalize their ML models at scale~\cite{Willem2020july}.  MLOps streamlines the end-to-end lifecycle of ML models by addressing aspects such as reliability, scalability, sustainability, and performance~\cite{BarredoArrieta2020Jun}.\\
The core components of MLOps encompass a variety of techniques, tools, and best practices that optimize the entire ML model lifecycle~\cite{Symeonidis2022}. The main principles of MLOps can be organized into four categories~\cite{Faubel_challenges}: 1) Data Engineering (Data Collection, Data Analysis, Data Preparation), 2) Model Engineering (Model Building, Model Training, Model Evaluation, Model Selection, Model Packaging), 3) Operations (CI/CD-testing, Model Deployment, Monitoring), 4) Supporting Activities (Infrastructure, Versioning, Automation, Tools).\\
MLOps inherited automation as a fundamental principle from DevOps. Employing continuous integration and continuous delivery (CI/CD) pipelines automates the various software development and deployment stages. This automation ensures that changes to code or data automatically trigger the deployment, leading to faster iterations and reliable releases. Furthermore, continuous training (CT) -- an additional practice in MLOps -- enables automatic model retraining, allowing models to remain up-to-date and adaptable to real-time data changes~\cite{Symeonidis2022}.\\
By embracing MLOps, organizations can effectively tackle challenges related to version control, reproducibility, model drift, data drift, and model performance degradation. Further, they can develop an end-to-end MLOps infrastructure considering the need for seamless explanation methods and leveraging explanations for the model tests, monitoring, improvement, and auditing. 

\subsection{MLOps Workflow}
\label{sec:related_word_mlops_workflow}

 Amershi, Saleema et al.'s paper \cite{Amershi2019May} discusses several crucial steps in the software engineering workflow for ML. 
 \begin{enumerate}
     \item First, the appropriate features are selected for a product in the requirements section.
     \item Then, the search for existing data sets and the acquisition of new data occurs, with incorrect data being cleaned from the data sets. For many ML methods, additional labeling is necessary.
     \item In the feature engineering stage, all activities for extracting and selecting features are carried out, and models are chosen and trained. If the features are not good enough, a new look is taken at them.
     \item In the model evaluation stage, metrics test the model with additional data sets.
     \item Finally, after the deployment of the model on the target platform, it is monitored.
 \end{enumerate}

MLOps, as described by Symeonidis et al. \cite{Symeonidis2022}, goes beyond this process by incorporating additional testing and continuous integration/continuous delivery (CI/CD) to ensure that ML is brought into operation smoothly and efficiently. Monitoring, sustainability, robustness, fairness, and explainability are core competencies for building mature, automated, and efficient MLOps systems. In their paper, Symeonidis et al. provide an overview of MLOps, defining the operation and the components of such systems while highlighting the current problems and trends. They also present different tools and their usefulness in providing the corresponding guidelines. Furthermore, they propose a connection between MLOps and AutoML (Automated Machine Learning), suggesting how this combination could work.



Personal data and the GDPR play a role in a few industrial applications. As per GDPR \cite{gdpr}, providing individuals with meaningful explanations is crucial when automated decisions are made \cite{Tamburri2020Sep}. To achieve this, MLOps and AI software sustainability are essential. However, the more these platforms are integrated into day-to-day software operations, the more the risk of AI software sustainability becoming unsustainable from a social, technical, or organizational perspective.

The challenges of operationalizing ML models in the manufacturing domain are significant, given the probabilistic nature of ML algorithms, reliance on large data sets, and the need for constant retraining \cite{Raffin2022Jan}. Raffin et al. \cite{Raffin2022Jan} have proposed a domain model which divides the landscape into five contexts reflecting the differences between edge systems, monitoring and dashboarding on the cloud instance, and the features of the MLOps domain. In their work, Raffin et al. \cite{Raffin2022Jan} refer to the white paper from Salama et al. \cite{salama2021practitioners}, which explains the overall MLOps process and the workflow needed per step in the MLOps process. The end-to-end workflow proposed in our paper has similar components to those presented in Raffin et al.'s work. However, it also shows the relationship between the components and intermediate artifacts, such as data sets, models, and serving packages. These critical process components need to incorporate explainable AI. Moreover, in industrial applications, some unique challenges need to be addressed. These involve particular challenges relevant to the process industry~\cite{Gartler2021Dec} and Industry~4.0~\cite{Faubel_challenges} context. 

\subsection{Explainable AI}
\label{related_XAI}
XAI describes that AI, especially ML solutions, should be understandable and explainable to stakeholders such as modelers and end users. At present, it describes a collection of different techniques and methods that attempt to achieve this goal \cite{project_proposal}. 
This is a response to the ``black box'' phenomena in ML, where even the designers cannot explain the reasoning behind a specific inference \cite{BarredoArrieta2020Jun}. XAI promises to help users perform more effectively by refining their understanding of AI-powered systems and dispelling misconceptions. XAI may also allow for the social right to explanation~\cite{gdpr}, although it is relevant even with no regulatory requirement. By improving the user experience of a product or service, XAI can help end users trust that the AI system is making good decisions. 
During the modeling process, stakeholders and users can assess the quality of the model and make necessary improvements to both the model and data, ultimately leading to better performance. Further, during model deployment, the explanation enables human oversight by helping the user judge whether a prediction is reasonable. XAI aims to make AI more transparent and understandable to humans, explain an AI inference in the past, present, and future, and reveal information based on actions. These characteristics make it possible to confirm existing knowledge, challenge it, and generate new assumptions \cite{project_proposal}.


Brennen \cite{Brennen2020apr} delves into the various interpretations of ``Explainable AI''. During his research, respondents had differing views on what they wanted to understand about AI and what they already understood. The need for explainability can arise from debugging, bias identification, and building trust in new technology. While decision trees are transparent by design, more opaque models such as deep neural networks (DNNs) require more complex monitoring and explanation \cite{Tamburri2020Sep}. Borg et al. \cite{Borg2021apr} suggest using heatmaps for visual explanations in computer vision models, and Dhanorkar et al. \cite{Dhanorkar2021jun} explore xai design space, including model selection and tracking adversarial model behaviors. Galhotra and pradhan \cite{Pradhan2022jun} categorize XAI methods as intrinsic or post hoc, while Cheng et al. \cite{Cheng2019may} discuss white-box vs.\ black-box and interactive vs.\ static explanations. In industrial IoT systems, time-series data is prevalent~\cite{giurgiu2019jan}, and LIME~\cite{ribeiro2016feb} and SHAP \cite{Lundberg2017dec} are used to explain univariate time-series classification algorithms~\cite{Mujkanovic2020jul}.

Further, transparency is important to ensure calibration of a user’s mental model to a system’s performance \cite{Dzindolet2003Jun,SantonideSio2018Feb} and overall ensure adequate human control \cite{Methnani2021Sep,SantonideSio2018Feb}. Hence, transparency is crucial to XAI \cite{BarredoArrieta2020Jun} since it ensures human control over the system's performance and good engineering practices \cite{Jobin2019Sep,trustworthy_ai_2023Jun}. 

\subsection{Interactive Machine Learning}
\label{sec:XIL}
Explainable Artificial Intelligence (XAI) aims to make ML results more interpretable, while Interactive Machine Learning (IML) involves integrating humans into the insight discovery process. Addressing the common obstacle of insufficiently labeled data in developing classification models for process monitoring and optimization in chemical batch production, particularly focusing on multivariate signal data, \cite{Ahmad2022Sep} propose an active learning web-application that assists human experts in labeling batch recipe steps using process data. To tackle the crucial task of dataset labeling in supervised and semi-supervised machine learning, \cite{Grimmeisen2020Dec} combines model-based active learning with user-based interactive labelling, to employ visual cues to guide users in selecting and labeling instances, leading to positive effects on user confidence, difficulty, orientation, and perception of model performance. \cite{Teso2019Jan,Schramowski2020Aug} highlights the increasing importance of IML and proposes Explanatory Interactive Learning (XIL) as a way to bridge the gap between XAI and IML. XIL combines algorithmic explanations with user interaction during iterative training loops, enabling users to adjust labels during the training process based on explanations and provided feedback which leads to enhancing the connection between XAI and IML. Assaf and Schumann \cite{ijcai2019p932} investigated CNN models for forecasting and explanations. These models were used to provide visual explanations, as \cite{Bernard2018Sep} introduced visual interactive labeling (VIAL), which combines active learning and interactive visualizations to leverage their respective strengths. In domains such as manufacturing, where high-dimensional data with limited labels and spurious correlations are prevalent, manual feature engineering can be expensive. To overcome this challenge, \cite{DimitraGkorou2020} propose a method called interactive visual feature engineering. By utilizing dimensionality reduction techniques and interactive visualizations. Application of XIL enhances the predictive capabilities and interpretability of models, empowering human experts in the process. The learning algorithm queries the user, predicts labels, and provides explanations, enabling iterative feedback and improvement of the model.

\section{Requirements for an Explainable MLOps Architecture}
\label{sec:requirements}

\outline{2}{Leila~Methnani, Thomas Woudsma, Leonhard}{
FOCUS!!!

Requirements:

- Stakeholders (Leila~Methnani)

- Summary of Requirements (Thomas Woudsma, Leonhard Faubel)

From:

- Brainstorming

- Interviews \cite{faubel_interviews}

- Other Documents

- Give one UC example (outcome), overview (Prodrive?)



}

We collected requirements for a novel MLOps software architecture that can support the EXPLAIN ML life cycle, reflecting the needs of various industrial domains like mining, paper pulp and metals production, power generation, and electronics manufacturing. 
These requirements are collected in multiple ways. For instance, interviews were conducted as part of a case study. Furthermore, in a brainstorming session with industrial experts, software engineers, and XAI researchers, we identified requirements regarding data collection and management, models, explainers, training, deployment and serving, monitoring and feedback, general architecture, infrastructure, and performance.

The results of this case study are published in~\cite{faubel_interviews}. This study aimed to investigate the extent to which MLOps is implemented by four project partners and describes their ML use cases, MLOps software architecture, tools, and requirements from their unique perspectives. Our interviews revealed that each industry partner uses MLOps differently, depending on their use case. There were variations in tools and architectural patterns used across the board. Overall, our findings were heavily focused on the architecture decisions involved in the MLOps tool landscape that the interviewed companies utilized.

Furthermore, a brainstorming workshop was held with the partners to gather more details about specific requirements for different domains and components of the architecture. As mentioned, the \ifcameraready EXPLAIN \fi \ifanonymous Anonymous \fi project covers a wide variety of use cases that can result in divergent requirements. Nonetheless, the goal is to identify overlapping requirements to find generic architecture components that can be reused for XAI applications in general. The requirements from this session have been split into MLOps requirements, e.g., for the infrastructure and storage, and XAI requirements, containing requirements, for instance, for the visualization of explanations and the connection between existing MLOps components and explainers. 
The requirements are also linked to both stakeholder groups, which are the domain experts during training, and end users during production. The requirements substructures are shown below:

\begin{description}
    \item[MLOps] Infrastructure, Data \& Storage, Data Traceability, Models, Model Traceability, Model Deployment \& Serving, Feedback, Monitoring, Other Non-functional
    \item[XAI] Explainer Support, Explainer Traceability, Explanation-based Review, Explainer Feedback, Explainer Monitoring
\end{description}


The MLOps requirements are listed in Section~\ref{sec:mlops_requirements} and the XAI requirements in Section~\ref{sec:xai_requirements}. In a requirement, \textit{shall} means that the application should fulfill this requirement and \textit{must} means that it must necessarily be fulfilled. As in the MLOps life cycle, for the stakeholder groups, a distinction is made between development and production requirements (D and P in Table~\ref{tab:MLOps_requirement_mapping}). 

The development steps are: 
\begin{itemize}
    \item \textbf{D1:} Requirement Identification
    \item \textbf{D2:} Data Collection
    \item \textbf{D3:} Explanatory Modeling
    \item \textbf{D4:} Statistical Validation
    \item \textbf{D5:} Explanation-based Review. 
\end{itemize} 

The production steps are: 
\begin{itemize}
    \item \textbf{P1:} Deploy
    \item \textbf{P2:} Load Live Data
    \item \textbf{P3:} Prepare Live Data
    \item \textbf{P4:} Output Explanations
    \item \textbf{P5:} Record model output, user and system response
    \item \textbf{P6:} Incremental Explanatory Training
    \item \textbf{P7:} Update Improved Model
\end{itemize}

\label{sec:requirement_elicitation}
\outline{}{}{
	Requirements from:
-	Brainstorming
o	Data collection and management
o	Models, explainers, training
o	Deployment and serving
o	Monitoring and feedback
o	General architecture, infrastruture, performance
-	Interviews
o	Conclusion
o	Reference to case study
-	Other documents: project proposal, WP2 collection and management, WP3 conventional ML approaches, WP4
}


\subsection{MLOps Requirements}
\label{sec:mlops_requirements}
Implementing MLOps requires careful consideration of divergent requirements, particularly those related to architecture. In this section, we will delve into the crucial MLOps requirements that are taken into account for our architecture and should form the basis of industrial ML applications. They are split into different categories and mapped to the life cycle phases (from Figure~\ref{fig:explain_MLOps_lifecycle}) in Table~\ref{tab:MLOps_requirement_mapping}. The next section extends these MLOps requirements with additional XAI requirements.

\vspace{-1em}
\begin{table}
    \caption{Mapping of the \textbf{MLOps} requirements to the different phases in the life cycle in Figure~\ref{fig:explain_MLOps_lifecycle}. The minor phases D1-D5 map to the five phases of the development cycle, and the minor phases P1-P7 map to the seven phases of the production cycle.}
    \label{tab:MLOps_requirement_mapping}
    \begin{tabularx}{\textwidth}{|X|c|c|c|c|c|c|c|c|c|c|c|c|}
        \hline
        \textbf{Major phases} & \multicolumn{5}{|c|}{\textbf{Development}} & \multicolumn{7}{|c|}{\textbf{Production}}\\
        \hline
        \emph{Minor phases} & D1 & D2 & D3 & D4 & D5 & P1 & P2 & P3 & P4 & P5 & P6 & P7\\
        \hline
        Infrastructure & X & X & X & X & X & X & X & X & X & X & X & X\\
        \hline
        Data \& Storage & & X & & & & & X & X & & & & \\
        \hline
        Data Traceability & & X & X & & & & X & X & & X & X & \\
        \hline
        Models & & & X & & & & & & & & &\\
        \hline
        Model Traceability & & & X & & & & & & & & &\\
        \hline
        Model Deployment \& Serving & & & & & X & X & & & X & X & X & X\\
        \hline
        Feedback & & X & X & & X & & & & & X & X &\\
        \hline
        Monitoring & & & & & & & & & & & X &\\
        \hline
        Other Non-functional & X & X & X & X & X & X & X & X & X & X & X & X\\
        \hline
    \end{tabularx}
\end{table}
\vspace{0em}

\subsubsection{Infrastructure (IN)}
In general, all components in the life cycle must adhere to the infrastructure requirements to guarantee that the platform is scalable \tag{MLOPS\-IN-\{05,07\}}, modular \tag{MLOPS-IN-\{01-04\}}, and maintainable \tag{MLOPS-IN-\{06\}}.
\begin{description}
    \item \tag{MLOPS-IN-01} The system must run on a cloud environment.
    \item \tag{MLOPS-IN-02} The system must run on an on-premises, self-hosted environment.
    \item \tag{MLOPS-IN-03} The system must support Windows and Linux applications.
    \item \tag{MLOPS-IN-04} The system must be composable to serve different UCs.
    \item \tag{MLOPS-IN-05} The system must support individual horizontal scaling for different components in the architecture.
    \item \tag{MLOPS-IN-06} The system should be described in a modeling language.
    \begin{description}
        \item \tag{MLOPS-IN-06.1} The system should be deployable with this modeling language.
        \item \tag{MLOPS-IN-06.2} The description must be version controlled.
    \end{description}
    \item \tag{MLOPS-IN-07} The system should support hardware acceleration for model training and inferences.
\end{description}

\subsubsection{Data \& Storage (DS)}
The system needs to support various use cases that deal with different data types \tag{MLOPS-DS-\{01\}}, storage \tag{MLOPS-DS-\{02,05\}}, and interfaces \tag{MLOPS-DS-\{03-04\}}. 
\begin{description}
    \item \tag{MLOPS-DS-01} The system must flexibly support different data formats, such as image, time series, and text data, e.g. \textit{CSV}, \textit{TXT}, and \textit{JSON}. 
    \item \tag{MLOPS-DS-02} The system must handle data storage sizes of Terabytes or more.
    \item \tag{MLOPS-DS-03} The system must be able to support data ingress via REST APIs and OPC UA.
    \item \tag{MLOPS-DS-04} The system must use open data standards and interfaces.
    \item \tag{MLOPS-DS-05} The system should support different backends such as SQL, NoSQL, InfluxDB, S3-compatible storage, and GCP buckets.
\end{description}

\subsubsection{Data Traceability (DT)}
The traceability of data is a critical functionality of any MLOPs system which deals lineage of data and metadata, including labels \tag{MLOPS-DT\-\{01\}}, the definition of data sets \tag{MLOPS-DT-02}, keeping track of applied preprocessing and generation \tag{MLOPS-DT-\{03\}}, and ingress of annotations \tag{MLOPS\-DT-\{04\}}. 
\begin{description}
    \item \tag{MLOPS-DT-01} The system must track the origin of the data.
    \begin{description}
        \item \tag{MLOPS-DT-01.1} This must track the equipment/hardware used, such as the sensors, including their location.
        \item \tag{MLOPS-DT-01.2} This must track when the data was captured with global timestamps (date and time with time zone).
        \item \tag{MLOPS-DT-01.3} This should track specific configurations, settings, and parameters of the sensor, e.g., the settings of a camera or the sampling frequency of a vibration sensor.
    \end{description}
    \item \tag{MLOPS-DT-02} The system must track the version of the data and data set, which are an immutable representation of the data at a certain point in time.
    \item \tag{MLOPS-DT-03} The system must track the preprocessing steps applied to the raw data before model training or inference.
    \begin{description}
        \item \tag{MLOPS-DT-03.1} The system should support saving transformed and artificially generated data that has been generated during training.
    \end{description}
    \item \tag{MLOPS-DT-04} The system must have an ingress for annotations and labels on the data.
    \begin{description}
        \item \tag{MLOPS-DT-04.1} The annotations must be linked to the samples.
    \end{description}
\end{description}

\subsubsection{Models (MO)}
Like with the data and storage requirements, due to the required support for different use cases, support different models and evaluations \tag{MLOPS-MO\-\{01-02\}} and interfaces \tag{MLOPS-MO\-\{03-04\}} is needed. 
\begin{description}
    \item \tag{MLOPS-MO-01} The system must support different ML frameworks such as TensorFlow, PyTorch, and scikit-learn models.
    \item \tag{MLOPS-MO-02} The system must support the following metrics: MAE, MSE, RMSE, F1 and $R^2$ score, Recall, Precision, and Specificity, Quantile Loss, and Variance Ratio Criterion.
    \item \tag{MLOps-MO-03} The system should support evaluation functions and aggregated metrics flexibly.
    \item \tag{MLOPS-MO-04} The system must support open model standards and interfaces.
\end{description}

\subsubsection{Model Traceability (MT)}
As with data traceability, model traceability is key for maintainable ML applications, with a link to the input of the model training \tag{MLOPS-MT-\{01,04\}}, the model architecture \tag{MLOPS-MT-\{02\}}, the software used for training \tag{MLOPS-MT-03}, and the final performance metrics \tag{MLOPS-MT-04}. 
\begin{description}
    \item \tag{MLOPS-MT-01} The system must track which data sets have been used for model training.
    \begin{description}
        \item \tag{MLOPS-MT-01.1} The system must track the training, validation, and test splits.
    \end{description}
    \item \tag{MLOPS-MT-02} The system must track the version of the model architecture used.
    \begin{description}
        \item \tag{MLOPS-MT-02.1} The system must track the initial state of the models before training.
    \end{description}
    \item \tag{MLOPS-MT-03} The system must track the software version used for training, including libraries, compilers, and interpreters, where applicable.
    \item \tag{MLOPS-MT-04} The system must track the performance metrics when the model is evaluated on a test data set.
    \item \tag{MLOPS-MT-05} The system must track the relationship between \tag{MLOPS-MO\-\{01-03\}}.
\end{description}

\subsubsection{Model Deployment \& Serving (MD)}
The next step is to deploy the models and serve incoming requests, where it is important that the models are versioned \tag{MLOPS-MD-01}, there are interfaces to retrain models and deploy models \tag{MLOPS\-MD-\{02-05\}}, and that all inference results are stored \tag{MLOPS-MD-\{06\}}. 
\begin{description}
    \item \tag{MLOPS-MD-01} The system must version the model artifacts after training (model registration).
    \item \tag{MLOPS-MD-02} The system should support automatically retraining models on triggers like new data sets or new annotations/labels.
    \item \tag{MLOPS-MD-03} The system should have an interface to start and define new model training runs manually.
    \item \tag{MLOPS-MD-04} The system should support different deployment schemes like shadow, canary, and A/B.
    \item \tag{MLOPS-MD-05} The system should have a GUI to deploy versioned model artifacts manually.
    \begin{description}
        \item \tag{MLOPS-MD-05.1} The GUI should show the deployed models, explainers, and deployment schemes.
    \end{description}
    \item \tag{MLOPS-MD-06} The system must support storing all inference results.
    \begin{description}
        \item \tag{MLOPS-MD-06.1} These predictions should be linked to the data the models used to generate them.
    \end{description}
\end{description}

\subsubsection{Feedback (FB)}
It is highly unlikely that the data on which the ML models make predictions are always of high quality or will not drift over time. Thus, it is important to have a proper data quality feedback interface \tag{MLOPS-FB-\{01\}} and ML prediction feedback and annotation interfaces \tag{MLOPS-FB-\{02-03\}}. 
\begin{description}    
    \item \tag{MLOPS-FB-01} The system must have an interface to view and provide feedback on the data quality.
    \begin{description}
        \item \tag{MLOPS-FB-01.1} This interface should allow marking data samples as 'bad', excluding them from being used by other components.
        \item \tag{MLOPS-FB-01.2} This interface must have the functionality to compare new data samples to similar data samples.
        \item \tag{MLOPS-FB-01.3} This interface should have the functionality to drill down on a sample and view its details and other related samples.
    \end{description}
    \item \tag{MLOPS-FB-02} The system must have a graphical user interface (GUI) to provide feedback on the predictions made by the models during training and in production.
    \begin{description}
        \item \tag{MLOPS-FB-02.1} This interface should have a way to guide users to the hard samples first, e.g., a sample that has an uncertain prediction or an incorrect prediction.
        \item \tag{MLOPS-FB-02.2} The system should have an interface to compare different training runs.
    \end{description}
    \item \tag{MLOPS-FB-03} The system should have a GUI for creating annotations and labels on the data.
    \begin{description}
        \item \tag{MLOPS-FB-03.1} This interface should support different data types, such as image and time series data.
        \item \tag{MLOPS-FB-03.2} This interface should support the different ML application types, such as classification and regression.
    \end{description}
\end{description}

\subsubsection{Monitoring (MT)}
Without monitoring the ML models' performance, it is unclear when feedback is required and when the performance of the data or models degrades. Requirements for ML model performance monitoring \tag{MLOPS-MT\-\{01,02\}}, data quality monitoring \tag{MLOPS-MT-\{03\}}, and system performance monitoring \tag{MLOPS-MT-03.3} are defined. 
\begin{description}
    \item \tag{MLOPS-MT-01} The system must have an interface that can show the actual performance metrics of the models in production.
    \item \tag{MLOPS-MT-02} The system should have an alerting system to indicate that feedback or retraining is required.
    \item \tag{MLOPS-MT-03} The system should have a component that measures the data quality, such as the drift over time or the changes in data with respect to a baseline.
    \begin{description}
        \item \tag{MLOPS-MT-03.1} The drift monitoring should have limits on these metrics to detect deterioration and send alerts.
        \item \tag{MLOPS-MT-03.2} This monitoring should work without providing any manual feedback on the predictions.
        \item \tag{MLOPS-MT-03.3} The system should monitor the latency and throughput of the components.
    \end{description}
\end{description}

\subsubsection{Other Non-functional (NF)}
For some use cases, an additional requirement is that the system comply with critical infrastructure regulations. Furthermore, open-source software is preferred to prevent vendor lock-in and create an architecture that is not dependent on closed-source, as long as it is maintained properly. 
\begin{description}
    \item \tag{MLOPS-NF-01} The system should comply if the applications need Safety and compliance regulations, such as the BSI KRITIS regulations ~\cite{noauthor_kritis_nodate}.
    \item \tag{MLOPS-NF-02} The system should use open-source packages that are maintained regularly and supported by a large community.
\end{description}

\subsection{XAI Requirements}
\label{sec:xai_requirements}

In this section, we will explore the key requirements for XAI. These requirements extend the existing MLOps practices and tools with additional explainer components. This means that most MLOps requirements must be implemented before any XAI requirements can be met. Table~\ref{tab:XAI_requirement_mapping} maps the XAI requirements to the life cycle phases of Figure~\ref{fig:explain_MLOps_lifecycle}.

\vspace{-1em}
\begin{table}
    \caption{Mapping of the \textbf{XAI} requirements to the different phases in the life cycle in Figure~\ref{fig:explain_MLOps_lifecycle}. The minor phases refer to the life cycle in the same way as in Table~\ref{tab:MLOps_requirement_mapping}.}
    \label{tab:XAI_requirement_mapping}
    \begin{tabularx}{\textwidth}{|X|c|c|c|c|c|c|c|c|c|c|c|c|}
        \hline
        \textbf{Major phases} & \multicolumn{5}{|c|}{\textbf{Development}} & \multicolumn{7}{|c|}{\textbf{Production}}\\
        \hline
        \emph{Minor phases} & D1 & D2 & D3 & D4 & D5 & P1 & P2 & P3 & P4 & P5 & P6 & P7\\
        \hline
        Explainer Support & & & X & & X & & & & X & & X & \\
        \hline
        Explainer Traceability & & & X & & X & & & & X & & X & \\
        \hline
        Explanation-based Review & & & X & & X & & & & & & & \\
        \hline
        Explainer Feedback & & & & & X & & & & & & X & \\
        \hline
        Explainer Monitoring & & & & & X & & & & & & X & \\
        \hline
    \end{tabularx}
\end{table}
\vspace{0em}

\subsubsection{Explainer Support (ES)}
These initial XAI requirements define the support for different explainer types and explanations \tag{XAI-ES-\{01,02\}} and optional data explainers \tag{XAI-ES-03}. 
\begin{description}
    \item \tag{XAI-ES-01} The system must support post-hoc explanations including feature attribution methods~\cite{swora_investigation_2023}. 
    \item \tag{XAI-ES-02} The system must support interpretable explanations, where the models either are interpretable themselves or provide explanations for their predictions~\cite{swora_investigation_2023}. 
    \item \tag{XAI-ES-03} The system may support data explanation methods, which provide insight into the underlying data structures without considering the predictions from the application models~\cite{swora_investigation_2023}. 
\end{description}

\subsubsection{Explainer Traceability (ET)}
Like traceability for the data and models, explainers also require traceability \tag{XAI-ET-01} to be able to reproduce results and improve them over time. They should be linked to the models and data as well \tag{XAI-ET-\{02-03\}}. 
\begin{description}
    \item \tag{XAI-ET-01} The system must track explainers used with each model.
    \item \tag{XAI-ET-02} The system must track generated explanations for different data and model combinations for use of feedback and review later.
    \item \tag{XAI-ET-03} The system should track which data is used to generate the explanations.
    \begin{description}
        \item \tag{XAI-ET-03.1} This should also track the domain knowledge used to generate explanations.
    \end{description}
\end{description}

\subsubsection{Explanation-based Review (ER)}
One of the major contributions of XAI to the life cycle is the explanation-based review phase, where explainers are integrated into existing review systems \tag{XAI-ER-01} and allow incremental improvements of the model during development \tag{XAI-ER-02}. 
\begin{description}
    \item \tag{XAI-ER-01} The system must integrate existing feedback interfaces to visualize explanations on the data used for training, validation, testing, and new data from production.
    \item \tag{XAI-ER-02} The system should support explanation-based reviewing, including eXplanatory Interactive (Machine) Learning (XIL)~\cite{Teso2019Jan}.
\end{description}

\subsubsection{Explainer Feedback (EF)}
Explainers are very useful in providing insights into the data and predictions from the models but also require feedback themselves \tag{XAI-EF-\{01,02\}} and visualization for different data and model combinations \tag{XAI-EF-\{03-05\}}.
\begin{description}
    \item \tag{XAI-EF-01} This system must have a GUI to provide feedback to the explainers.
    \begin{description}
        \item \tag{XAI-EF-01.1} This system should be integrated with existing feedback interfaces.
    \end{description}
    \item \tag{XAI-EF-03} The system must have a GUI to compare multiple explainers and models with the same data.
    \item \tag{XAI-EF-04} The system must have a GUI to compare different data with the same model and explainer.
    \begin{description}
        \item \tag{XAI-EF-04.1} This interface should be composable to add new explanations or visualizations for different applications.
    \end{description}
\end{description}

\subsubsection{Explainer Monitoring (EM)}
Finally, like with ML models, the performance of explainers should also be monitored to verify that they are still providing the right explanations for the right reasons.
\begin{description}
    \item \tag{XAI-EM-01} The system should have a component that tracks the performance of the explainers.
    \begin{description}
        \item \tag{XAI-EM-01.1} This should include explanation completeness, e.g., how much features contribute to the final model output.
        \item \tag{XAI-EM-01.2} This should include explanation stability, e.g., how much slight changes influence the model in the input data.
        \item \tag{XAI-EM-01.3} This should include explanation fidelity, e.g., how much the explanation approximates the prediction of the original complex model.
        \item \tag{XAI-EM-01.4} This should include explanation relevance, e.g., how much irrelevant information is not shown.
    \end{description}
\end{description}

\section{Architecture}
\label{sec:results}

\outline{2}{Mohsen~Nosratinia and Thomas Woudsma, Leonhard}{

- what is the motivation for the explainability architecture: what makes the architecture special (in the sense of MLOps and especially of expandability, where do we need explainability for what (see proposal)

briefly describe the architecture description (abstractly)

- describe categories/domains: Data Management, Model Training, Model Serving, Model Monitoring, User Feedback (Leonhard, proof read: Thomas, Mohsen)

- describe components (Leonhard, proof read: Mohsen~Nosratinia and Thomas Woudsma)

- integrated architectural view missing: but excerpts/ parts of it are already realized/tested by the partners: which experiences did we make?



}

\begin{figure}[ht]
    \includegraphics[width=\textwidth]{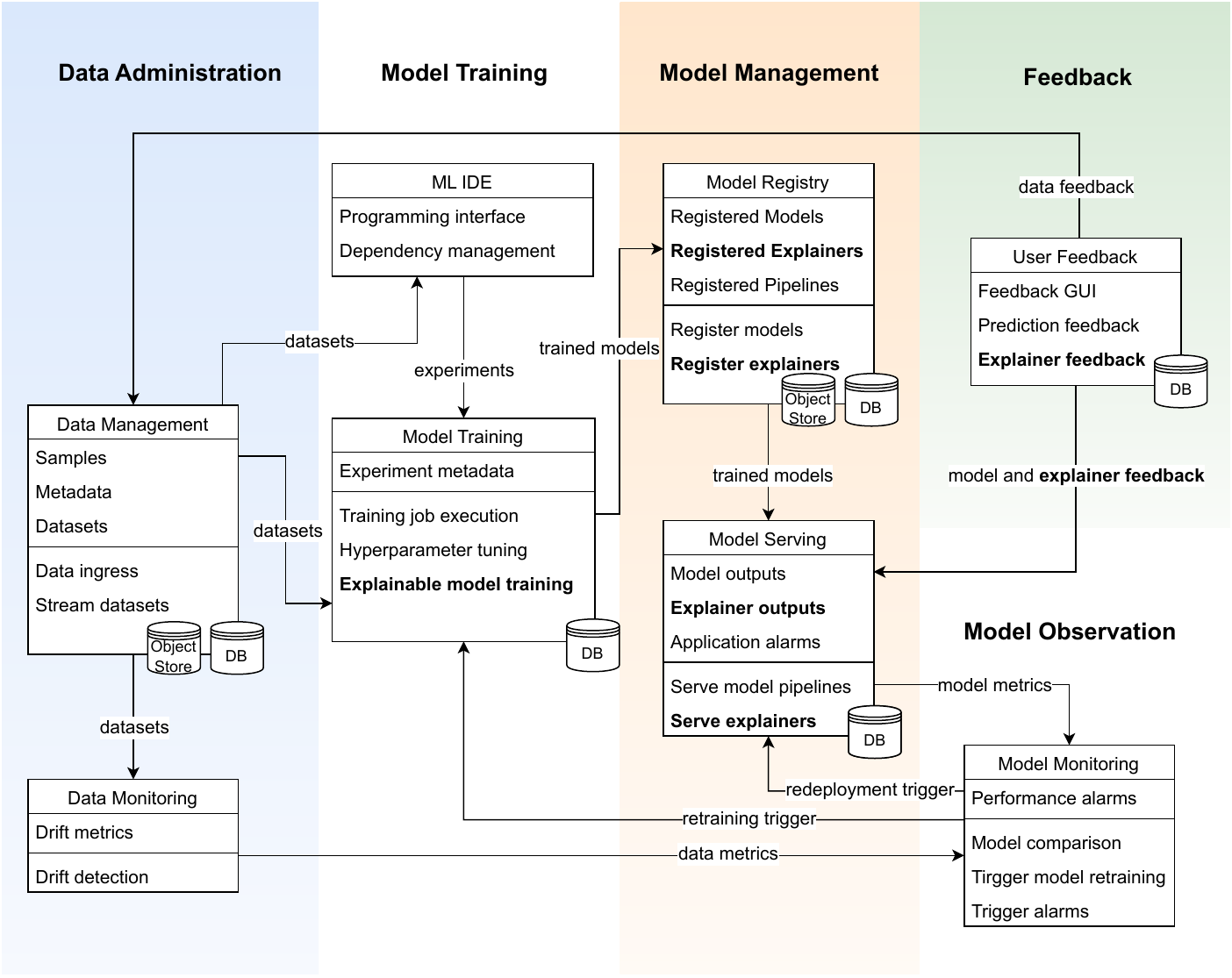}
    \caption{\ifcameraready EXPLAIN \fi \ifanonymous Anonymous \fi MLOps Software Architecture in five domains and eight components. Major XAI functionality in \textbf{bold}, but it is not limited to only those components.} \label{fig:architecture}
\end{figure}

In this chapter, we propose our MLOps software architecture that integrates explanation methods flexibly. 
As shown in the MLOps life cycle in Figure \ref{fig:explain_MLOps_lifecycle}, the explainable life cycle involves tasks from different domains that need to be carried out. It is essential to follow the life cycle to ensure the process is transparent and understandable to the stakeholders involved. For this reason, the architectural components are categorized according to the corresponding domains: Data Administration, Model Training, Model Management, Feedback, and Model Observation. The architecture is presented in Figure~\ref{fig:architecture}. The components partially use object stores and databases; arrows show the flow of data, e.g., data sets are used by the ML IDE and the model training component, or triggers, e.g., model monitoring triggers retraining of models. 
The architecture domains are Described in Section~\ref{sec:arch_domains} while the components are described in Section~\ref{sec:arch_components}. 

\subsection{Domains}
\label{sec:arch_domains}

The steps of our MLOps life-cycle are covered in the architecture simplified by the generic terms of our architecture domains: \textit{1.~data administration}, \textit{2.~model training}, \textit{3.~model management}, \textit{4.~user feedback}, and \textit{model observation}. 
\begin{enumerate}
    \item \textbf{\textit{Data administration (blue)}} includes the steps \textit{requirements identification}, \textit{data collection}, \textit{load live data} and \textit{prepare live data} from Figure~\ref{fig:explain_MLOps_lifecycle}. 
    
    \item \textbf{\textit{Model training (white)}} consists of \textit{interactive training} and \textit{update improved model}. 
    
    \item \textbf{\textit{Model management (orange)}} is reflected by \textit{deploy}.
    
    \item Finally, the \textbf{\textit{user feedback} and \textbf{\textit{model observation (green)}}} include \textit{record model output, user, and system response}. 
\end{enumerate}

    The components that reflect the explanatory functions of the MLOps life-cycle are shown together with their relation to \textit{explanatory modeling}, \textit{explanatory review}, \textit{output explanations}, and \textit{incremental explanatory training} in Table~\ref{tab:components in task domains}. 


\begin{table}
    \caption{Overview of the components and their relation to the explainability steps. An~X means that an explainability step is partially or completely covered by a component, an (X) means that the component depends on the result of a step.}\label{tab:components in task domains}
    \begin{tabularx}{\textwidth}{|X|c|c|c|c|}
        \hline
        \textbf{Domain}/Component & \textbf{Explanatory} & \textbf{Explanatory} & \textbf{Output}    & \textbf{Incremental}\\
                                &    \textbf{Modeling} &    \textbf{Review}  & \textbf{Explanations}  &  \textbf{Training} \\
        \hline
        \textbf{Data Administration} & & & &\\
        Data Management &  X & X & &\\
        Data Monitoring  & & & (X) &\\\hline
        \textbf{Model Training} & & & &\\
        ML IDE Component & X & X & &\\
        Training Component & X & X & & X\\\hline
        \textbf{Model Management} & & & &\\
        Model Registry & & X & &\\
        Model Serving & X & & X &\\\hline
        \textbf{Feedback} & & & &\\
        User Feedback & X & & X & X \\\hline        
        \textbf{Model Observation} & & & &\\
        Model Monitoring & & & (X) & X\\\hline
    \end{tabularx}
\end{table}

\subsection{Components}
\label{sec:arch_components}
\begin{enumerate}
    \item \textbf{Data Administration}
    \begin{description}
        \item[Data Management] The data management component encapsulates functionality to collect, annotate, and version data. All other relevant metadata is also collected to trace each data sample to its source fully. Next, it defines data sets, which are immutable representations of data and metadata to be used by other components in the architecture. They form the basis of reproducibility. The data can be stored in a database that suits the managed data samples, for example, S3-compatible image storage. The data monitoring component accesses data using data management to check whether the data distributions match the expectations. It is also used by the user feedback component, which can access the data for visualization, after which user feedback can be written back into the data management component.
        \item[Data Monitoring] The model monitoring component is dedicated to tracking the performance of the models and explainers. It monitors data and concept drifts and has the functionality to detect this kind of distribution shift in the data automatically. The model monitoring component also uses these data metrics, which can give a holistic view of the performance of the ML applications this architecture serves. Our architecture provides dashboards for end users, including alarms and notifications when models or explainers are not functioning correctly. 
        User feedback can be important to detect loss in performance - especially for tasks like anomaly detection, where no label is readily available to check for correctness. In cases of underperformance, our architecture can trigger model retraining automatically if the performance is below a certain threshold.  \\
    \end{description}
    
    \item \textbf{Model Training}
    \begin{description}
        \item[ML IDE] The ML IDE component enables data scientists and ML engineers to conduct initial experiments within the model training domain. This component can also retrieve data sets from the data management component so it is clear with which immutable data the experiments were run. Together with the data set identifier, all other training parameters, such as the software, model architecture, and hyperparameters, are logged to the model training component. The ML IDE component does not standardize a specific way of working but can be the basis for new ML models and pipelines (data processing steps and model executions).  This component serves as a platform to run training jobs on a cluster, which can speed up training. Further, the ML IDE is where hyperparameter tuning can be initialized. 

        \item[Model Training] The model training component enables the execution of existing training pipelines and hyperparameter tuning automatically. The data sets required for training the models and, potentially, explainers are streamed from the data management component. Lastly, the model training component must track all experiments. This includes the used data sets, training architecture and pipelines, hyperparameters, and software used. This component also has a view that allows for comparing and selecting the best-trained models using the hyperparameters and metrics as selectors.  \\
    \end{description}

    \item \textbf{Model Management}
    \begin{description}
        \item[Model Registry] The model registry component can register pipelines and take advantage of versioning capabilities, including explainers. This will easily integrate predictions into a production environment, making the process more efficient and effective. 
        \item[Model Serving] The platform's model serving component is designed to be scalable and can accommodate multiple registered and deployed models. Once deployed, the models can serve requests from other production services. The platform also keeps track of the inputs and outputs of the models and explainers it serves. This helps seamlessly update models to new versions e.g., using shadow or canary serving schemes, ensuring a smooth user experience. \\
    \end{description}

    \item \textbf{Feedback}
    \begin{description}
        \item[User Feedback] The user feedback component is a valuable tool for operators to have control over the ML process. It allows them to provide feedback on the functionality and quality of a model and also enables them to label data. This feature is crucial to ensure the model performs accurately and effectively. As an operator, having the ability to intervene in the ML process and provide feedback is essential to achieving the desired outcomes. It is also the component that visualizes the explainers in application-specific dashboards, ensuring operators can correctly interpret the predictions made by the models.\\
    \end{description}
    
    \item \textbf{Model Observation} 
    \begin{description}
        \item[Model Monitoring] The model monitoring component keeps track of a ML model's performance over time. It constantly evaluates the model's accuracy and identifies any declines in performance that may occur. Monitoring the model can quickly detect any issues with the data or the model itself and send alerts to the users. This allows for prompt action to rectify the situation. This helps to maintain the model's accuracy and usefulness over time. Next, it also allows for comparing different models on the data to get an idea of which one performs better. Finally, it can also trigger automatic retraining of existing models when there are new data and annotations for the model to use and then deploy the new models after they have been registered.
        \\
    \end{description}

\end{enumerate}



\section{Current State}
\label{sec:discussion}

We still need to implement the architecture described fully, but we have already implemented some components for testing and improving the architecture. In the majority of cases, partners use their existing components and extend them to implement the architecture. However, these are intended for internal use only and not for publication. The \ifcameraready University of Hildesheim \fi \ifanonymous ANONYMOUS \fi started by implementing a solution that even external parties can access \cite{gitlab_initial_architecture}. Since not all the explanation methods are available in current libraries and tools, this solution includes manuals and several example XAI implementations. 
Here, \textit{MLflow} can store and track explanations. The explanation methods can be stored and versioned using common code repositories. However, the feedback component still is in a prototypical stage and does not yet allow for feedback on explanations or interactions with explanation methods.  
The components used for our general solution are summarized in Table~\ref{tab:components in general solution}. 

\begin{table}
    \caption{Current components of the general solution \cite{gitlab_initial_architecture}.}\label{tab:components in general solution}
    \begin{tabularx}{\textwidth}{|X|X|}
        \hline
        \textbf{Component} & \textbf{Implementation} \\
        \hline
        Data Management         & Apache NiFi, MLflow \\
        Data Monitoring         & Grafana, Evidently \\
        Model Training          & Kubernetes, MLflow \\
        ML IDE                  & Jupyter, ML workspace \\
        Model Management        & MLflow \\
        Explainability Tools    & H20, MLflow, IBM Watson AI explainability 360 \\
        Model Serving           & Seldon.core, MLflow.deployment, Kubeflow \\
        Model Monitoring        & Evidently, River, Grafana \\
        Feedback Component      & Not yet implemented / Prototypical state \\
        \hline
    \end{tabularx}
\end{table}



Next, we describe an example implementation of a software system for managing image data in the electronics domain. The system includes a self-developed feedback \textit{Django} GUI for labeling and correcting predictions, a scalable training platform, a model registry, and model deployment capabilities. For model training, a \textit{Ray} cluster is being set up, to allow for a scalable training platform. The model training experiments are tracked with \textit{MLflow}, which is also used as a model registry. 
The components are deployed on a self-hosted, on-premises \textit{Kubernetes} cluster, except for the monitoring component, which uses a \textit{Splunk} dashboard (performance monitoring). 
Several image classification models are deployed with \textit{Seldon Core} which allows for horizontal scalability of model deployments in \textit{Kubernetes}. 

The electronics use case team is actively improving the feedback GUI for XAI with new explainer visualization capabilities. They plan to incorporate a state-of-the-art library of explainers designed explicitly for image data. The explainers will be registered in the model registry and served through the model serving component, allowing other components to utilize and benefit from them as well. This also requires the explainers to be registered (in the model registry) and served (in the model serving component), so other components can use them. Yet, the software currently lacks automated retraining, alerting, and data monitoring capabilities. 


\section{Conclusion}
\label{sec:conclusion}

\outline{-1.5}{Leonhard + All}{

- Conclusion

}

The increasing popularity of ML in industrial operations has opened up a range of opportunities for businesses to optimize their processes and improve productivity. However, deploying and managing ML models in production environments can be complex and challenging, especially in industries requiring high levels of explainability and transparency.

To address these challenges, we developed a novel MLOps software architecture that provides an integrated approach to MLOps, allowing for the integration of explanations into the development and deployment processes. Several industrial companies and universities have adopted this architecture, which will be validated and improved in the future within the project \ifcameraready EXPLAIN\fi \ifanonymous Anonymous\fi.
One of the key benefits of this architecture is its ability to support the explainability of ML models in industrial operations. This is particularly important in industries where regulatory requirements or ethical considerations demand high transparency and accountability.
Integrating explanations into the development and deployment processes also enables businesses to understand the behavior of their ML models better and identify potential issues or biases. This can improve the accuracy and reliability of the models and increase trust and confidence in their outputs.

Overall, developing this novel MLOps software architecture is a significant step forward in integrating XAI into industrial operations. In the future, we plan to evaluate our implementation and gain experience with the explainability components. We will work towards making it more user-friendly and easier to understand. For this reason, we will develop instructions, solutions, and example implementations, if necessary, to facilitate their use. This way, users will better understand the system and be able to use it more efficiently. As technology evolves, it will be interesting to see how businesses leverage these new capabilities to optimize their processes, improve efficiency, and drive innovation. 

\ifcameraready
\section*{Acknowledgment}
This work is supported by the project EXPLAIN, funded by the Federal Ministry of Education under grant 01—S22030E, funded by the Netherlands Enterprise Agency RVO under grant AI212001, and funded by Sweden´s Innovation Agency (Vinnova) under grant 2021-04336. Any opinions expressed herein are solely by the authors and not the funding agencies. 
\fi

\bibliographystyle{splncs04.bst}
\bibliography{bibliography.bib}

%




\end{document}